\documentclass[aps,prl,twocolumn,floatfix,showpacs]{revtex4-1}
\usepackage{graphicx}
\usepackage{amsfonts,amsmath,amssymb}
\usepackage{amsthm}
\usepackage{dsfont,bm}
\usepackage{color,soul}
\usepackage{xcolor, soul}
\sethlcolor{yellow}
\usepackage{amsbsy}
\usepackage[colorlinks=true,linkcolor=blue,pagecolor=blue,filecolor=blue,menucolor=blue,urlcolor=blue,citecolor=blue,anchorcolor=blue]{hyperref}%

\usepackage{mathbbol}

\usepackage{sidecap}

\begin{document}

\title{Transverse Spin current at the normal/p-wave magnet junctions}

\author{Ali Akbar Hedayati}
\affiliation{Department of Physics, University of Tehran, Tehran, Iran}
\author{Morteza Salehi}
\affiliation{Department of Physics, Bu-Ali Sina University, Hamadan, Iran}

\date{\today}

\begin{abstract}
We investigate the transmission properties of a junction between a normal metal and a p-wave magnet in the ballistic regime. We introduce a two-dimensional square lattice that confirms the p-wave magnet criteria. The $\alpha$-vector of the magnet breaks the inversion symmetry of the parabolic dispersion, shifting them in $k$-space. These shifts alter the propagation direction of fermions passing through the junction interface. Depending on spin orientation, the transmission process exhibits anisotropic, angle-dependent behavior. We also demonstrate a mirror symmetry between fermions with opposite spin directions, leading to the emergence of a transverse spin current that flows parallel to the interface. Additionally, we show that the $\alpha$-vector acts as a source for the dynamics of the spin-density wave and observe the formation of an indirect gap in the conductance of the junction. Our findings highlight the unique transmission characteristics and spin transport phenomena in normal metal/p-wave magnet junctions, paving the way for potential applications in spintronic devices.

\end{abstract}

\maketitle
\section{Introduction}
Producing spin current is one of the most important areas of spintronics\cite{Zutic2004RMP} that can be obtained via different methods, such as ferromagnet/semiconductor spin injection or spin pumping\cite{Tserkovnyak2005RMP}. The spin injection occurs via the manipulation of itinerant spin imbalance in the ferromagnet lead \cite{Tsymbal2011Book}. On the other hand, in the spin pumping the spin current can be created in the adjacent materials via spin imbalance of the magnetic materials such as antiferromagnets or ferromagnets\cite{Tserkovnyak2002PRB,Tserkovnyak2002PRL,Xiao2014PRL,Takei2015PRB}. 

Altermagnets represent a novel class of magnetic materials that exhibit unique spin textures and transport properties, distinguishing them from traditional ferromagnets and antiferromagnets\cite{Bai,Mazin2022PRX,Yan2024APL}. Unlike ferromagnets, which possess a uniform spin alignment, and antiferromagnets, characterized by alternating spin orientations \cite{Stefanita2012Book}, altermagnets exhibit spin-momentum locking and no net magnetization\cite{Liu2022PRX}. These materials are classified by spin-group symmetries\cite{Smejkar2022PRX}. The crystal rotation or inversion symmetry ($\mathcal{P}$) relates the opposite spins of different sublattices whereas the time-reversal symmetry ($\mathcal{T}$) is broken. This effect lifts the spin degeneracy  and leads to anisotropic bands in the $k$-space such as \cite{Hirsch1990PRB,Krempasky2024Nature}:
\begin{equation}
\left\lbrace
 \begin{array}{l}
E_\sigma(\textbf{k})=E_\sigma(-\textbf{k})\\
E_\sigma(\textbf{k})\neq E_{-\sigma}(\textbf{k})
\end{array}\right.
\label{Eq.d-wave symmetry}.
\end{equation}
Here, $\boldsymbol{\sigma}$s are Pauli matrices that act on the spin degree of freedom. So far, the classification was confined into even-parity wave (d, g or i-wave) altermagnets \cite{Smejkal2022PRX2,Smejkal2020SciAdv,Fedchenko2024SciAdv}. Even-parity altermagnets are predicted theoretically and confirmed experimentally in various compounds such as $Mn_5Si_3$\cite{Smejkal2022PRX3,Reichlova2024NC}, $CrSb$\cite{Guo2023MPT,Ding2024arXiv}, $FeSb_2$\cite{Mazin2021PNAS},  $MnTe$\cite{Mazin2023PRB,Krempasky2024Nature,Lee2024PRL,Osumi2024PRB,Orlova2024ArXiv} and $RuO_2$ \cite{Ahn2019PRB,Berlijn2017PRL,Liao2024PRL}. Altermagnets attract lots of attention because of its interesting features and future applications\cite{Ang2023arXiv,Sun2023PRB,Ouassou2023PRL,Beenakker2023PRB,Hodt2024PRB,Soori2024PRB,Soori2023JPCM,Papaj2023PRB,Banerjee2024PRB}. 

\begin{figure}
\includegraphics*[scale=0.13]{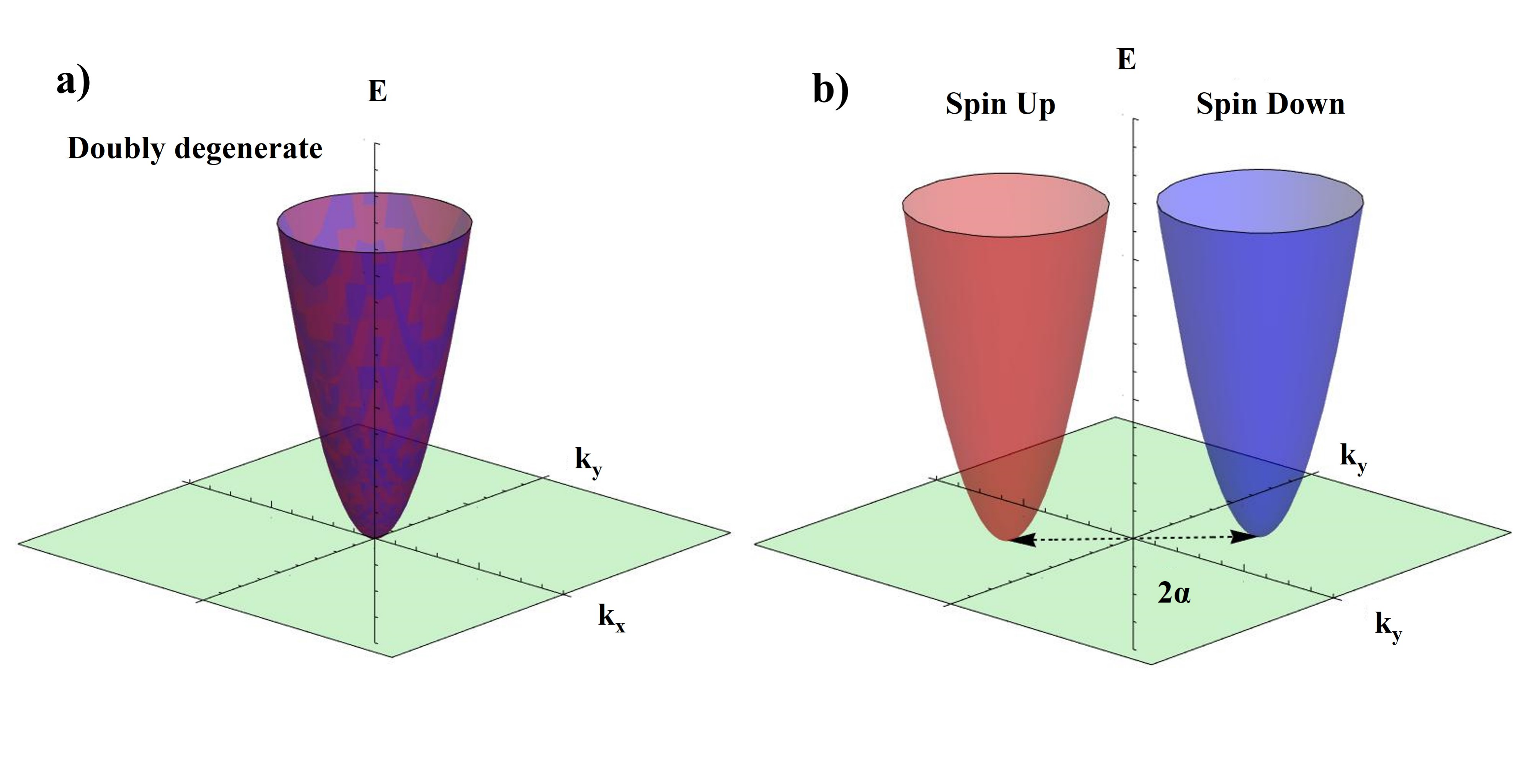}
\caption{a) The doubly degenerate dispersion of normal state that is located in the $\Gamma$-point of the Brillouin zone. b) In the presence of $\alpha$-vector, the $\mathcal{P}$-symmetry has been broken. The Kramer's degeneracy of spin polarized bands has been lifted by their's horizontal shift with respect to the origin. }
\label{Fig.k-dispersion}
\end{figure}

There is other class, dubbed p-wave magnet, which has two types of collinear and non-collinear odd-parity spin-texture in the momentum space. In this category, the $\mathcal{T}$-symmetry is preserved while the $\mathcal{P}$-symmetry is broken\cite{hellenes2024pwavemagnets},

\begin{equation}
\left\lbrace 
\begin{array}{l}
E_\sigma(\textbf{k}) \neq E_\sigma(-\textbf{k})\\
E_\sigma(\textbf{k})= E_{-\sigma}(\textbf{-k})
\end{array}
 \right. 
 \label{Eq.P-waveSymmetry}.
\end{equation}
Our focus is in the collinear odd-parity type, where phase- and spin-dependent hoppings on a single square lattice break $\mathcal{P}$-symmetry, leading to p-wave magnetization. The non-collinear odd-parity magnet with coplanar spin texture that leads to compensated magnetic phase is disscussed recently\cite{hellenes2024pwavemagnets}.
The proposed candidate of non-collinear p-wave magnet is $CeNiAsO$. In this compound, the $C_{2\bot}$ symmetry is broken while the $\mathcal{T}$ is preserved, where $C_{2\bot}$ is the rotation along the collinear magnetization axes. As schematically shown in Fig.(\ref{Fig.k-dispersion}), these symmetries impose horizontal shift on the spin-polarized bands and create even nodes in the magnetization texture to satisfy the condition of  Eq.(\ref{Eq.P-waveSymmetry}) \cite{hellenes2024pwavemagnets}.

\begin{figure}
\includegraphics*[scale=0.25]{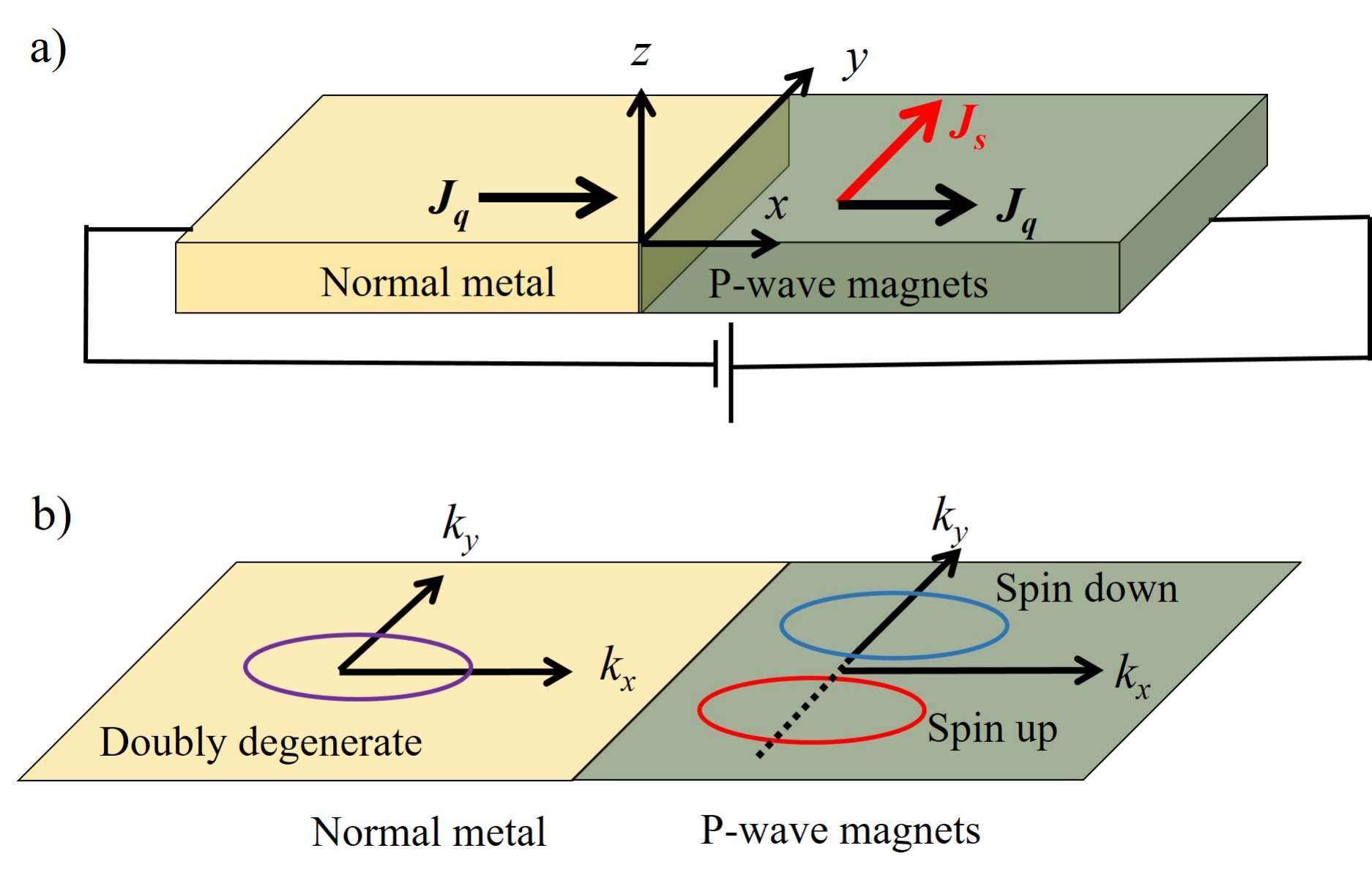}
\caption{a) A real space junction of normal/p-wave magnet. In the presence of the differential gate voltage, charge current, $J_q$, flows in the x-direction whereas the transverse spin current, $J_s$, in the p-wave magnet side of the junction flows parallel to the y-direction. b) A $k$-space version of the normal/p-wave magnet junction, the center of degenerate Fermi circle is located at the $\Gamma$-point of the $k$-space whereas the center of spin-polarized Fermi circles are moved according to the $\boldsymbol{\alpha}$ vector. }
\label{Fig.junction}
\end{figure}

In this work, we present a tight-binding model on a square lattice with phase- and spin-dependent hopping to the nearest neighbors, which breaks $\mathcal{P}$-symmetry while preserving $\mathcal{T}$-symmetry. With our model, we obtain the low-energy effective Hamiltonian near the high-symmetry points of the Brillouin zone, consistent with its phenomenological description. As shown in Fig. (\ref{Fig.junction}), we consider a junction between a normal metal and a p-wave magnet in the ballistic limit, investigating how the energy band's shift influences the transport properties of the junction. We find that the junction interface bends the propagation direction of incoming fermions based on their spin orientation in opposite directions. Consequently, the transmission probability across the junction depends on both the spin and propagation angle of the incoming fermions. This effect manifests in a mirror symmetry that generates a transverse spin current. Additionally, we explore the spin density wave (SDW) and demonstrate that the $\alpha$-vector can act as a source in its continuity equation.

\section{Theory and Formalism}

\begin{figure}
\includegraphics*[scale=0.1]{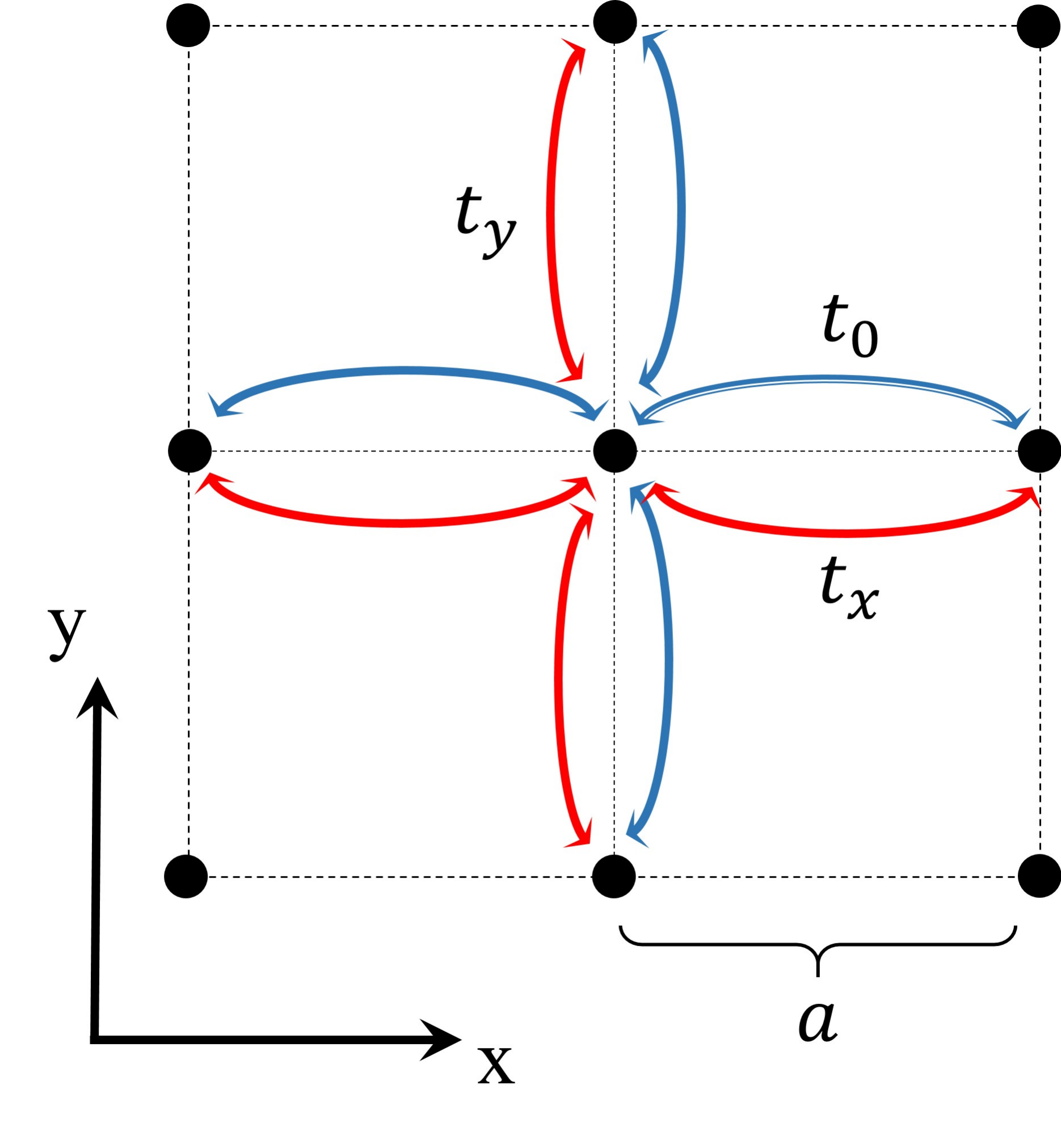}
\caption{A minimal tight-binding model for p-wave magnets. The lattice constant of the square lattice is $a$. The usual hopping processes between nearest neighbors are shown by the $t_0$ term (blue arrows). The essence of p-wave magnet orders can be obtained by phase- and  spin-dependent hopping processes between the nearest neighbors. These effects are shown by ${ t_x, t_y}$ terms (red arrows), respectively.}
\label{Fig.MinimalModel}
\end{figure}

\subsection{Tight-binding Hamiltonian}
The phenomenological description of the low-energy effective Hamiltonian for p-wave magnets that shifts horizontally the spin-polarized bands in the $k$-space is given by\cite{Maeda2024Arxiv,Bernevig2006PRL},
\begin{equation}
\begin{array}{rl}
H(\textbf{k}) & =  \frac{\hbar^2}{2m}\left( (\textbf{k}^2+\boldsymbol{\alpha}^2) \sigma_0 + 2 \textbf{k}.\boldsymbol{\alpha} \sigma_z\right)-\mu \\
& \\
& = \left(\begin{array}{cc}
H_\uparrow(\textbf{k}) & 0 \\
0 & H_\downarrow(\textbf{k}) \\
\end{array}\right)
\end{array}.
\label{Eq.H_Tanaka}
\end{equation}

Here, m is the effective mass and $\hbar$ is the reduced Planck constant. Also, $\boldsymbol{\alpha}=(\alpha_x, \alpha_y)$ is the magnet strength vector and $\mu$ acts as the chemical potential. The spin-polarized Hamiltonians are given by
\begin{equation}
H_{\uparrow(\downarrow)}(\textbf{k})=\frac{\hbar^2}{2m}\left( \textbf{k}\pm \boldsymbol{\alpha} \right)^2-\mu.
\end{equation}

To understand the physics of a p-wave magnet, we construct a tight-binding model. As shown in Fig.(\ref{Fig.MinimalModel}), one can consider a square lattice with a constant of $a$. The $t_0$ is the usual hopping between the nearest neighbors that is shown by the blue arrows. The p-wave magnet can be achieved by phase- and spin-dependent hopping to the nearest neighbors that are shown by red arrows. This Hamiltonian can be written as below,

\begin{eqnarray}
\mathcal{H}_{lattice}=\frac{-t_0}{2}\sum_{j_x,j_y,\delta}\left(\hat{C}_{j_x,\delta}^{\dagger}\hat{C}_{j_x+a,\delta}+\hat{C}_{j_y,\delta}^{\dagger}\hat{C}_{j_y+a,\delta}+H.C\right) \nonumber\\
 +\frac{ t_x}{2}\sum_{j_x,j_y,\delta,\gamma}\left(e^{i\phi}\hat{C}_{j_x,\delta}^{\dagger}\sigma^z_{\delta,\gamma}\hat{C}_{j_x+a,\gamma}+e^{-i\phi}\hat{C}_{j_x+a,\gamma}^{\dagger}\sigma^z_{\delta,\gamma}\hat{C}_{j_x,\delta}\right)\nonumber \\
 +\frac{ t_y}{2}\sum_{j_x,j_y,\delta,\gamma}\left(e^{i\phi}\hat{C}_{j_y,\delta}^{\dagger}\sigma^z_{\delta,\gamma}\hat{C}_{j_y+a,\gamma}+e^{-i\phi}\hat{C}_{j_y+a,\gamma}^{\dagger}\sigma^z_{\delta,\gamma}\hat{C}_{j_y,\delta}\right)\nonumber
 \\
 \label{Eq.Hamiltonian_Real_SQ}.
\end{eqnarray}
Where ${\delta, \gamma}={\uparrow, \downarrow}$ determine the spin configuration of creation, $\hat{C}{j}^\dagger$, or annihilation, $\hat{C}_j$, operations on the $\textbf{j}$-site of the lattice. The phase of $\phi$ breaks the $\mathcal{P}$-symmetry of the lattice and $\sigma^z_{\delta,\gamma}$ is the z-component of the Pauli spin matrix that tunes the spin-dependent hopping between nearest neighbors\cite{Tohid2024}. We assume the strength of hopping can be different in perpendicular directions. Also, the phase-dependent hopping can be achieved via a magnetic vector potential\cite{Marder2010Book}. In the case of $\phi=\pi/2$, the $\mathcal{T}$-symmetry of the system is preserved and the conditions of Eq.(\ref{Eq.P-waveSymmetry}) are satisfied. It is straightforward to obtain the Fourier transform of this Hamiltonian in the $k$-space as,
\begin{equation}
\mathbf{H}=\sum_{\textbf{k}}\hat{C}^\dagger_\textbf{k}\mathcal{H}(\textbf{k})\hat{C}_\textbf{k}.
\end{equation}
Here, $\hat{C}^\dagger_\textbf{k}=\left(C^\dagger_{\textbf{k},\uparrow},C^\dagger_{\textbf{k},\downarrow} \right)$ and $\hat{C}_\textbf{k}=\left(C_{\textbf{k},\uparrow},C_{\textbf{k},\downarrow} \right)^T$ are creation and annihilation operators in the $k$-space, respectively. For the Hamiltonian in the $k$-space we have,

\begin{equation}
\begin{array}{rl}
\mathcal{H}(\textbf{k})=& -t_0\left(cos(k_x a)+cos(k_y a)\right)\sigma_0 \\
& \\
& + \left(t_x sin(k_x a)+t_y sin(k_y a)\right)\sigma_z
\end{array}.
\label{Eq.Hamiltonian_k_space}
\end{equation}
The conduction and the valence bands touch each other at the $\Gamma$- and $M$- points of the Brillouin zone. To obtain the low-energy effective Hamiltonian, one can expand Eq.(\ref{Eq.Hamiltonian_k_space}) near the $\Gamma$- point. With these substitutions,
\begin{equation}
\begin{array}{l}
\alpha_x=t_x/t_0 a \\
\\
\alpha_y=t_y/t_0 a \\
\\
m=t_0 a^2/\hbar^2 \\
\\
\mu= (4t_0^2+t_x^2+t_y^2)/2t_0\\
\end{array}
\label{Eq.Substitution}
\end{equation}
Eq.(\ref{Eq.Hamiltonian_k_space}) reduces to Eq.(\ref{Eq.H_Tanaka}). As shown schematically in Fig.(\ref{Fig.k-dispersion}), the amplitude of $t_x$ and $t_y$ move the spin-polarized bands horizontally around the origin. In the next subsection, we explore the effect of the band's shift on the transport properties.

\begin{figure}
\includegraphics*[scale=0.22]{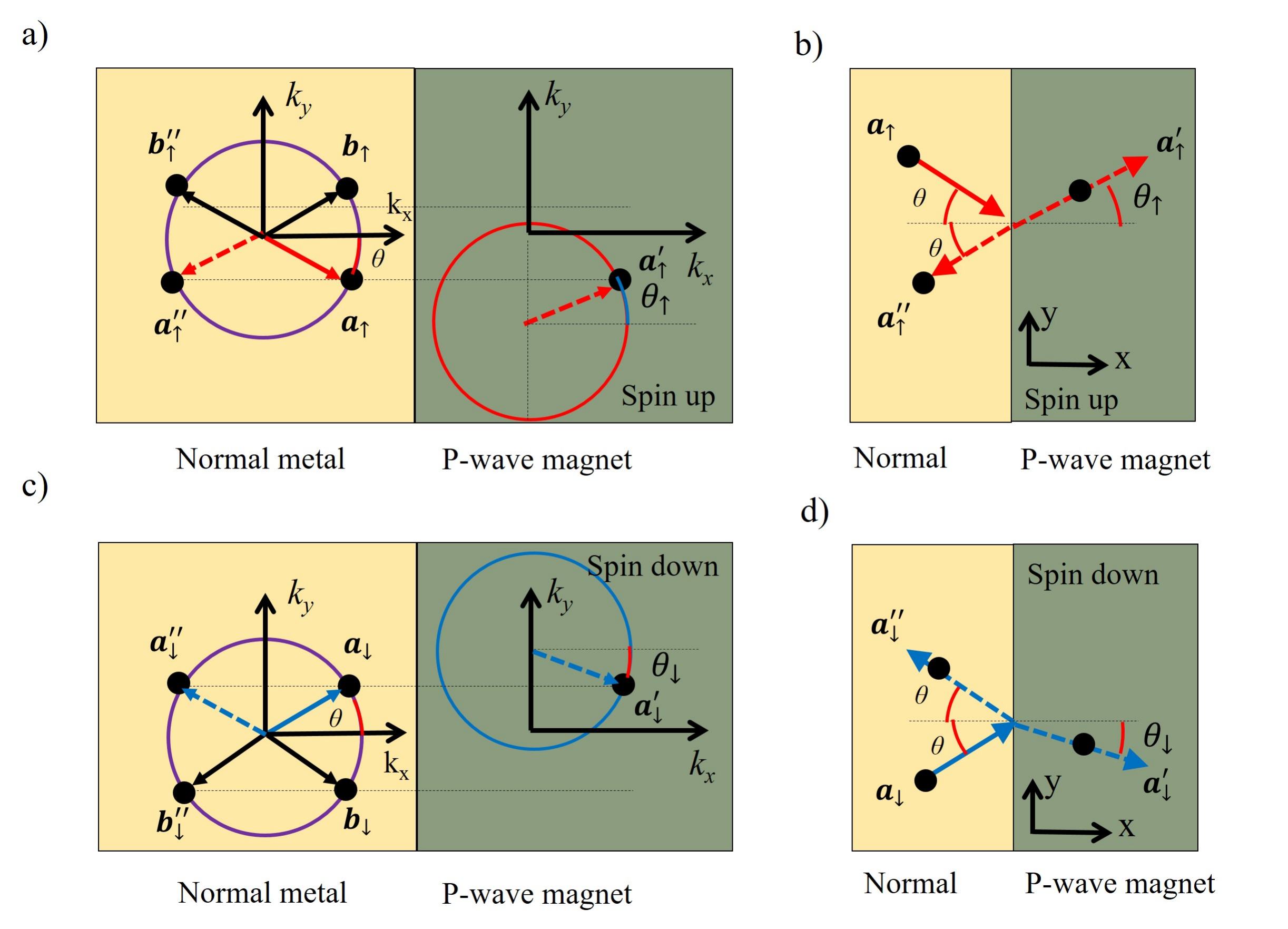}
\caption{a) A top view of the normal/p-wave magnet junction in the $k$-space for the spin-$\uparrow$ sub-band. The origin of the spin$\uparrow$ sub-band is located at the center in the normal side while its origin moves according to the $\boldsymbol{\alpha}$-vector on the other side of the junction. Here, we assume ${\alpha_x = 0, \alpha_y \neq 0}$. In the ballistic limit, the $k_y$ component of the wave vector is conserved during the scattering. So, the fermion in the $a_\uparrow$ of the Fermi circle has a chance to transport across the interface and find a stable state on the other side of the junction in $a'\uparrow$. Also, there is a probability of R for $a\uparrow$ to reflect from the interface and sit on the $a"\uparrow$ of the related Fermi circle. The red and black arrows that originate from the center of the Fermi circles determine the propagation direction of their related states in real space. In this scenario, the $b\uparrow$ cannot pass through the interface and is reflected to $b"\uparrow$ with the probability of $R=1$. b) The scenario of part (a) is shown in real space, where the incoming fermion from the $a\uparrow$ state at the angle of $\theta$, passes through the interface into the $a'\uparrow$ with the probability of T and propagates at the angle of $\theta\uparrow$. c) A similar procedure is followed for the spin-$\downarrow$ sub-band for the normal/p-wave magnet junction in $k$-space. d) This shows the real-space version of the scattering processes in part (c).}
\label{Fig.4}
\end{figure}

\subsection{The bending of propagation direction }
\label{S.Bending}
We consider a junction between a normal metal and p-wave magnet in the ballistic limit. The real-space demonstration is shown in part (a) of Fig.(\ref{Fig.junction}), and its $k$-space version is depicted in its part (b). To investigate the transport properties we use Eq.(\ref{Eq.H_Tanaka}) that satisfies the p-wave magnet condition of Eq.(\ref{Eq.P-waveSymmetry}). Since $\boldsymbol{\alpha}=0$, we have Kramers' degeneracy in the normal side of the junction, $x \leq 0$. So, the origin of both bands are located at the center of $k$-space. In the normal region, the eigenvalues of Eq.(\ref{Eq.H_Tanaka}) are,    

\begin{equation}
    E_{\uparrow (\downarrow)}=\frac{\hbar^2}{2m}(k_x^2+k_y^2)-\mu ,
    \label{Eq.eigenvalues_normal}
\end{equation}
 where, $\{k_x, k_y\}$ are the components of the wave vector. There is a Fermi circle for each energy with the radius of $\sqrt{2m(E+\mu)/\hbar^2}$ that demonstrates the stable and available states. On the normal side, this Fermi circle is shown in Figs.(\ref{Fig.junction},\ref{Fig.4}) in purple color. Since the value of the wave vector can be obtained by $  k=\sqrt{k_x^2+k_y^2}$, we can define $\theta$ as the propagation angle of fermions in the normal side,
\begin{equation}
    \begin{array}{l}
    k_{x}=k \cos\theta \\
    k_{y}=k \sin\theta 
    \end{array}
    \label{Eq.k_N}.
\end{equation}
The group velocity $v_{x(y)}=\partial E/\hbar \partial k_{x(y)}$ confirms that $\theta$ is the propagation angle of the states in the real-space. This is shown by the vectors centered at the origin of the Fermi circles in Fig.(\ref{Fig.4}). With a standard procedure, the wave functions of Eq.(\ref{Eq.eigenvalues_normal}) can be obtained as,
\begin{equation}
    \psi^\pm_\uparrow(x,y)=\frac{1}{\sqrt{v_x}}\left(
    \begin{array}{c}
         1 \\
         0 
    \end{array}
    \right) e^{\pm ik_x x+i k_y y},
    \label{Eq.Up_normal_WF}
\end{equation}
\begin{equation}
    \psi^\pm_\downarrow(x,y)=\frac{1}{\sqrt{v_x}}\left(
    \begin{array}{c}
        0 \\
         1 
    \end{array}
    \right) e^{\pm ik_x x+i k_y y}.
    \label{Eq.Down_normal_WF}
\end{equation}
Here, the $\pm$ signs indicate the propagation direction of fermions in real space. On the p-wave magnet region ($ 0 \leq x $), where the $\boldsymbol{\alpha} \neq 0$, the Kramers' degeneracy is broken and the eigenvalues of Eq.(\ref{Eq.H_Tanaka}) can be written as,

\begin{equation}
\begin{array}{l}
    E'_\uparrow=\frac{\hbar^2}{2m}\left( (k'_{x,\uparrow}+\alpha_x)^2+(k'_{y,\uparrow}+\alpha_y)^2\right)-\mu\\
    \\
    
    E'_\downarrow=\frac{\hbar^2}{2m}\left( (k'_{x,\downarrow}-\alpha_x)^2+(k'_{y,\downarrow}-\alpha_y)^2\right)-\mu\\  
\end{array}
\label{Eq.eigenvalues_altermagnet}
\end{equation}
Here the prime indicates the p-wave magnet side of the junction. In the same energy, ($E'_\uparrow=E'_\downarrow=E'$) , Eq.(\ref{Eq.eigenvalues_altermagnet}) shows two separated Fermi circles with the same radius of $\sqrt{2m(E'+\mu)/\hbar^2}$. The related dispersions of Eq.(\ref{Eq.eigenvalues_normal}) and Eq.(\ref{Eq.eigenvalues_altermagnet}) are depicted in Fig.(\ref{Fig.k-dispersion}). Similar to the normal region, we define the wave vector as $K'_\uparrow=\sqrt{K'^2_{x,\uparrow}+K'^2_{y,\uparrow}}$, where $\{K'_{x,\uparrow}=k'_{x,\uparrow}+\alpha_x, K'_{y,\uparrow}=k'_{y,\uparrow}+\alpha_y\}$. Now, one can use the propagation angle for the spin-$\uparrow$ sub-band, $\theta'_\uparrow$, to derive the x-component of the group velocity as,
\begin{equation}
v'_{x,\uparrow} = \frac{\hbar}{m} K'_\uparrow \cos\theta'_\uparrow
\end{equation}
For the spin-$\downarrow$ sub-band, there is a similar procedure.
\begin{equation}
v'_{x,\downarrow} = \frac{\hbar}{m} K'_\downarrow \cos\theta'_\downarrow
\end{equation}
The related wave functions of Eq.(\ref{Eq.eigenvalues_altermagnet}) can be written as,
\begin{equation}
    \psi^\pm_\uparrow(x,y)=\frac{1}{\sqrt{v'_{x,\uparrow}}}\left(
    \begin{array}{c}
         1 \\
         0 
    \end{array}
    \right) e^{\pm ik'_{x,\uparrow} x+i k'_{y,\uparrow} y},
    \label{Eq.Up_Alter_WF}
\end{equation}
\begin{equation}
    \psi^\pm_\downarrow(x,y)=\frac{1}{\sqrt{v'_{x,\downarrow}}}\left(
    \begin{array}{c}
        0 \\
         1 
    \end{array}
    \right) e^{\pm ik'_{x,\downarrow} x+i k'_{y,\downarrow} y}.
    \label{Eq.Down_Alter_WF}
\end{equation}
Since Eq.(\ref{Eq.H_Tanaka}) is diagonal in the spin space, there is no spin-flipped process during the scattering from the interface.
In the ballistic limit, the energy and the parallel component of wave vectors are conserved during the scattering processes. 
\begin{equation}
\left\{
\begin{array}{l}
E=E'_\uparrow=E'_\downarrow , \\
\\
k_y=k'_{y,\uparrow}=k'_{y,\downarrow}.
\end{array}
\right.
\label{Eq.Ballistic_Conditions}
\end{equation}
In a normal/p-wave magnet junction, the first condition of Eq.(\ref{Eq.Ballistic_Conditions}) forces the radius of  all Fermi circles to be equal to each other. The second condition bends the propagation direction of transported fermions. To illustrate its influence on the scattering processes , we assume $\{\alpha_x=0, \alpha_y\neq 0 \}$. In this case, the Fermi circles moves in the $k_y$-direction oppositely.  As shown in Part(a) of Fig.(\ref{Fig.4}), the incoming fermion with a propagation angle of $\theta$ is located on the $a_\uparrow$ state of the degenerate Fermi circle. This fermion hits the interface from the normal side of the junction and has a probability of $T_\uparrow(E,\theta)$ to transport across the junction. Also, there is a probability of $R_\uparrow(E,\theta)$ to be reflected from the interface. To satisfy the second condition of Eq.(\ref{Eq.Ballistic_Conditions}), this fermion must be located into the $a'_\uparrow$ state for the transported case or into the $a"_\uparrow$ state for the reflected case. Due to the difference between $\theta$ and $\theta_\uparrow$, there is a bend in the propagation direction of transported fermions. This process occurs when two Fermi circles of the same spin sub-bands from both sides of the junction overlap along the $k_y$-direction. Outside the overlap region, the $b_\uparrow$ state, there is no probability for the incoming fermion to find a stable state on the other side of the junction. In this case the hitting fermion must be reflected into $b"_\uparrow$ state with the probability of $R=1$. This indicates that the interface of the junction is sensitive to the propagation direction of hitting fermions anisotropically. In part (b) of Fig.(\ref{Fig.4}), the bending procedure is illustrated in real-space. The same scenario that occurs for the spin-$\downarrow$ sub-band, are shown in parts (c) and (d) of Fig.(\ref{Fig.4}). There is a very important difference in the spin-$\downarrow$ scattering process. The bending of propagation direction occurs opposite to the spin-$\uparrow$ sub-band. This leads to a spin current that flows parallel to the interface. 

\subsection{Transport properties}
To calculate the transport probability, we construct the wave function of spin-$\uparrow$ sub-band in the normal region,
\begin{equation}
\psi_\uparrow(x,y)=\psi_\uparrow^+(x,y)+r_\uparrow \psi^-_\uparrow(x,y),
\label{Eq.WFinN}
\end{equation}
where $r_\uparrow$ is the reflection amplitude. For the p-wave magnet region, we have
\begin{equation}
\psi'_\uparrow(x,y)=t_\uparrow \psi'^+_\uparrow(x,y),
\label{Eq.WFinAM}
\end{equation}
where $t_\uparrow$ is the transmission amplitude. Using the boundary conditions at $x=0$\cite{Maeda2024Arxiv},
\begin{equation}
\begin{array}{ll}
\psi_\uparrow(0,y)=\psi'_\uparrow(0,y) \\
\\
v_x\psi_\uparrow(0_-,y)-v'_{x,\uparrow}\psi'_\uparrow(0_+,y)=\frac{2}{i\hbar}U_0 \psi'_\uparrow(0_+,y)
\end{array},
\label{Eq.BC}
\end{equation}
one can calculate the probability amplitudes of $r_\uparrow$ and $t_\uparrow$. Here, $U_0$ is the interface potential that can exist in real situations. After some straightforward algebra, we derive the probability of transmission as,
\begin{equation}
T_\uparrow(E,\theta)=t_\uparrow t^*_\uparrow=\frac{4 cos\theta \cos\theta'_\uparrow}{(\cos\theta+\cos\theta'_\uparrow)^2+U_0^2}
\label{Eq.T_up}
\end{equation}
For spin-$\downarrow$ sub-band, the $T_\downarrow(E,\theta)$ can be obtained in a similar way. Since the energy and $k_y$ are conserved, we can use them to find the overlap region of the Fermi circles. From the Eq.(\ref{Eq.eigenvalues_altermagnet}), the $k'_{x,\uparrow}$ and $k'_{x,\downarrow}$ are derived as below,

\begin{equation}
\begin{array}{l}
      k'_{x,\uparrow}=\pm \sqrt{\frac{2m}{\hbar^2}(E+\mu)-(k_y+\alpha_y)^2}-\alpha_x \\
      \\
      k'_{x,\downarrow}=\pm \sqrt{\frac{2m}{\hbar^2}(E+\mu)-(k_y-\alpha_y)^2}+\alpha_x
\end{array}
\label{Eq.kx_AM}
\end{equation}
The wave functions of Eq(\ref{Eq.Up_Alter_WF}) and Eq.(\ref{Eq.Down_Alter_WF}) are in propagating mode when the square root function of Eq.(\ref{Eq.kx_AM}) be positive. The negative sign of the square root function leads to evanescent modes that do not contribute to the transport. For the spin-$\uparrow$ sub-band, the overlap of the Fermi circles can be obtained via the allowed propagation angle,
\begin{equation}
\sin^{-1}(\max\{-1,-1-\alpha_y/k\})\leq \theta \leq \sin^{-1}(\min\{1,1-\alpha_y/k\})
\label{Eq.Spin_up_allowed}.
\end{equation}
In a similar way, the allowed propagation angle of spin-$\downarrow$ sub-band can be calculated as,
\begin{equation}
\sin^{-1}(\max\{-1,-1+\alpha_y/k\})\leq \theta \leq \sin^{-1}(\min\{1,1+\alpha_y/k\})
\label{Eq.Spin_down_allowed}.
\end{equation}
For $k < |\alpha_y|/2$, the Fermi circles on both sides of the junction do not overlap in $k_y$-direction and all incoming particles reflect from the interface. It means an indirect gap appears in the junction whereas the energy dispersions of all bands on the both sides of the junction are gapless.

\subsection{The origin of spin current}
In the Sec.(\ref{S.Bending}), we use the ballistic conditions of Eq.(\ref{Eq.Ballistic_Conditions}) to demonstrate that a transverse spin current flows parallel to the interface. In the Landau-Lifshitz-Gilbert (LLG) theory\cite{Tsymbal2011Book}, the appearance of spin-polarized current is accompanied by a source like spin-transfer torque in the steady state approximation. Here, we show that $\boldsymbol{\alpha}$-vector acts as a source in the continuity equation of the spin density wave and produces the transverse spin current.

The second quantization form of Eq.(\ref{Eq.H_Tanaka}) can be written in real space as follows,
\begin{equation}
\begin{array}{rl}
\mathcal{H}_{R} & =\int d\textbf{r} \hat{\Psi}^\dagger(\textbf{r})H(\frac{\hbar}{i}\nabla_\textbf{r})\hat{\Psi}(\textbf{r}) \\
& \\
& =\int d\textbf{r} \hat{\Psi}^\dagger_\uparrow(\textbf{r})H_\uparrow(\frac{\hbar}{i}\nabla_\textbf{r})\hat{\Psi}_\uparrow(\textbf{r}) \\
& \\
& +\int d\textbf{r} \hat{\Psi}^\dagger_\downarrow(\textbf{r})H_\downarrow(\frac{\hbar}{i}\nabla_\textbf{r})\hat{\Psi}_\downarrow(\textbf{r})
\end{array}
.
\label{Eq.H_R}
\end{equation}
 
 Here, the $R$ index stands for real space. Also, $\hat{\Psi}^\dagger(\textbf{r})=(\hat{\Psi}^\dagger_\uparrow(\textbf{r}), \hat{\Psi}^\dagger_\downarrow(\textbf{r}))$ and $\hat{\Psi}(\textbf{r})=(\hat{\Psi}_\uparrow(\textbf{r}), \hat{\Psi}_\downarrow(\textbf{r}))^T$ are the creation and annihilation field operators. These operators obey the anti-commutation relation and Fermi-Dirac statistics. Also, their commutations with $\mathcal{H}_R$ are,
 \begin{equation}
 \begin{array}{l}
 \left[H_\uparrow(\frac{\hbar}{i}\nabla_\textbf{r}), \hat{\Psi}^\dagger_\uparrow(\textbf{r}) \right]=\frac{1}{2m}\left(\frac{\hbar}{i}\nabla_\textbf{r}-\boldsymbol{\alpha}\right)^2 \hat{\Psi}^\dagger_\uparrow(\textbf{r}),\\
 \\
 \left[H_\uparrow(\frac{\hbar}{i}\nabla_\textbf{r}), \hat{\Psi}_\uparrow(\textbf{r}) \right]=\frac{-1}{2m}\left(\frac{\hbar}{i}\nabla_\textbf{r}+\boldsymbol{\alpha}\right)^2 \hat{\Psi}_\uparrow(\textbf{r})
 \end{array}.
\label{Eq.Commutation_Psi}
 \end{equation}
 There are similar relations for $\hat{\Psi}_\downarrow(\textbf{r})$ and $\hat{\Psi}^\dagger_\downarrow(\textbf{r})$. Due to the diagonal form of Eq.(\ref{Eq.H_Tanaka}), the density operator can be decomposed into the spin-polarized density as,
 
 \begin{equation}
 \begin{array}{rl}
 \rho(\textbf{r}) & = \hat{\Psi}^\dagger(\textbf{r})\hat{\Psi}(\textbf{r}) \\
 &\\
 & = \hat{\Psi}^\dagger_\uparrow(\textbf{r})\hat{\Psi}_\uparrow(\textbf{r})+\hat{\Psi}^\dagger_\downarrow(\textbf{r})\hat{\Psi}_\downarrow(\textbf{r})\\
 & \\
 & = \rho_\uparrow(\textbf{r})+\rho_\downarrow(\textbf{r})\\
 \end{array}
 \label{Eq.ProbabilityDensity}
 \end{equation}
 
 The dynamics of any time-independent operator, $\hat{A}$, can be obtained by means of the Heisenberg equation of motion,
 $\partial_t \hat{A}=i[H,\hat{A}]/\hbar$. This can be used to calculate the dynamics of the density as,
 \begin{equation}
 \begin{array}{rl}
 \partial_t \rho(\textbf{r}) & = \partial_t \rho_\uparrow(\textbf{r})+\partial_t \rho_\downarrow(\textbf{r})\\
 & \\
 & = (\partial_t \hat{\Psi}^\dagger_\uparrow(\textbf{r}))\hat{\Psi}_\uparrow(\textbf{r})+ \hat{\Psi}^\dagger_\uparrow(\textbf{r})(\partial_t\hat{\Psi}_\uparrow(\textbf{r}))\\
 & \\
 &  + (\partial_t \hat{\Psi}^\dagger_\downarrow(\textbf{r}))\hat{\Psi}_\downarrow(\textbf{r})+ \hat{\Psi}^\dagger_\downarrow(\textbf{r})(\partial_t\hat{\Psi}_\downarrow(\textbf{r}))\\
 \label{Eq.Paritial_rho}
 \end{array}
 \end{equation}
This leads to the continuity equation, $\partial_t \rho(\textbf{r})+\nabla_\textbf{r}.J(\textbf{r})=0$. Here, the density current,  $J(\textbf{r})=J_\uparrow(\textbf{r})+J_\downarrow(\textbf{r})$, is decomposed into the spin-polarized density current,

\begin{equation}
\begin{array}{ll}
\textbf{J}_{\uparrow}(\textbf{r}) & =\frac{\hbar}{2im}\left(\hat{\Psi}^\dagger_\uparrow(\textbf{r})(\nabla_{\textbf{r}}\hat{\Psi}_\uparrow(\textbf{r}))- (\nabla_\textbf{r}\hat{\Psi}^{\dagger}_\uparrow(\textbf{r}))\hat{\Psi}_\uparrow(\textbf{r}) \right) \\
& \\
&+\frac{\boldsymbol{\alpha}}{m}\rho_{\uparrow}(\textbf{r})\\
& \\
& = \textbf{J}^0_\uparrow(\textbf{r})+\frac{\boldsymbol{\alpha}}{m}\rho_{\uparrow}(\textbf{r})
\end{array},
\label{Eq.Current_Density_Up}
\end{equation}
and, 
\begin{equation}
\begin{array}{ll}
\textbf{J}_{\downarrow}(\textbf{r}) & =\frac{\hbar}{2im}\left(\hat{\Psi}^\dagger_\downarrow(\textbf{r})(\nabla_{\textbf{r}}\hat{\Psi}_\downarrow(\textbf{r}))- (\nabla_\textbf{r}\hat{\Psi}^{\dagger}_\downarrow(\textbf{r}))\hat{\Psi}_\downarrow(\textbf{r}) \right) \\
& \\
&-\frac{\boldsymbol{\alpha}}{m}\rho_{\downarrow}(\textbf{r})\\
& \\ 
& = \textbf{J}^0_\downarrow(\textbf{r})-\frac{\boldsymbol{\alpha}}{m}\rho_{\downarrow}(\textbf{r})
\end{array}.
\label{Eq.Current_Density_Up}
\end{equation}
In the absence of an imbalance between the spin-polarized density, $\delta\rho(\textbf{r})=\rho_{\uparrow}(\textbf{r})-\rho_\downarrow(\textbf{r})=0$, the current density is independent of the magnet strength vector. Also, the current density is conserved, $\textbf{J}(\textbf{r})=\textbf{J}^0(\textbf{r})=\textbf{J}^0_\uparrow(\textbf{r})+\textbf{J}^0_\downarrow(\textbf{r}).$

Since the collinear magnetization is parallel to the $z$-axis, we define the spin density operator as below,
\begin{equation}
\begin{array}{rl}
\hat{S}_z & =\hat{\Psi}^\dagger(\textbf{r})\sigma_z \hat{\Psi}(\textbf{r}) \\
& \\
& = \hat{\Psi}^\dagger_\uparrow(\textbf{r}) \hat{\Psi}_\uparrow(\textbf{r})-\hat{\Psi}^\dagger_\downarrow(\textbf{r}) \hat{\Psi}_\downarrow(\textbf{r}) \\
& \\
& = \rho_{\uparrow}(\textbf{r})-\rho_{\downarrow}(\textbf{r})
\end{array}
\label{Eq.SDW}
\end{equation}
Similar to Eq.(\ref{Eq.Paritial_rho}), one can derive the continuity equation for spin-density wave such as,
\begin{equation}
 \begin{array}{rl}
 \partial_t \hat{S}_z(\textbf{r}) & = \partial_t \rho_\uparrow(\textbf{r})-\partial_t \rho_\downarrow(\textbf{r})\\
 & \\
 & = (\partial_t \hat{\Psi}^\dagger_\uparrow(\textbf{r}))\hat{\Psi}_\uparrow(\textbf{r})+ \hat{\Psi}^\dagger_\uparrow(\textbf{r})(\partial_t\hat{\Psi}_\uparrow(\textbf{r}))\\
 & \\
 &  + (\partial_t \hat{\Psi}^\dagger_\downarrow(\textbf{r}))\hat{\Psi}_\downarrow(\textbf{r})+ \hat{\Psi}^\dagger_\downarrow(\textbf{r})(\partial_t\hat{\Psi}_\downarrow(\textbf{r}))\\
 \label{Eq.Paritial_S_z}
 \end{array}.
\end{equation}
One can use Eq.(\ref{Eq.Commutation_Psi}) to obtain the continuity equation,

\begin{equation}
\partial_t \hat{S}_z+\nabla_\textbf{r}.\textbf{J}^s(\textbf{r})=\frac{-\boldsymbol{\alpha}}{m}.\nabla_\textbf{r}\rho(\textbf{r}),
\label{Eq.Continuity_Sz}
\end{equation}
where the spin-polarized current is defined as, $\textbf{J}^s(\textbf{r})=J^0_\uparrow(\textbf{r})-J^0_\downarrow(r)$.  The Eq.(\ref{Eq.Continuity_Sz}) manifests the role of magnet strength vector as a source for the continuity equation of spin-density wave. In the special case of Fig.(\ref{Fig.4}), the spin current that flows parallel to the interface has $J^0_\uparrow(\textbf{r})=-J^0_\downarrow(\textbf{r})$. It means there is no charge current flows along the $y-$direction whereas the spin-polarized current becomes a pure spin current.
\section{Results and discussion}

\begin{figure}
\includegraphics*[scale=0.55]{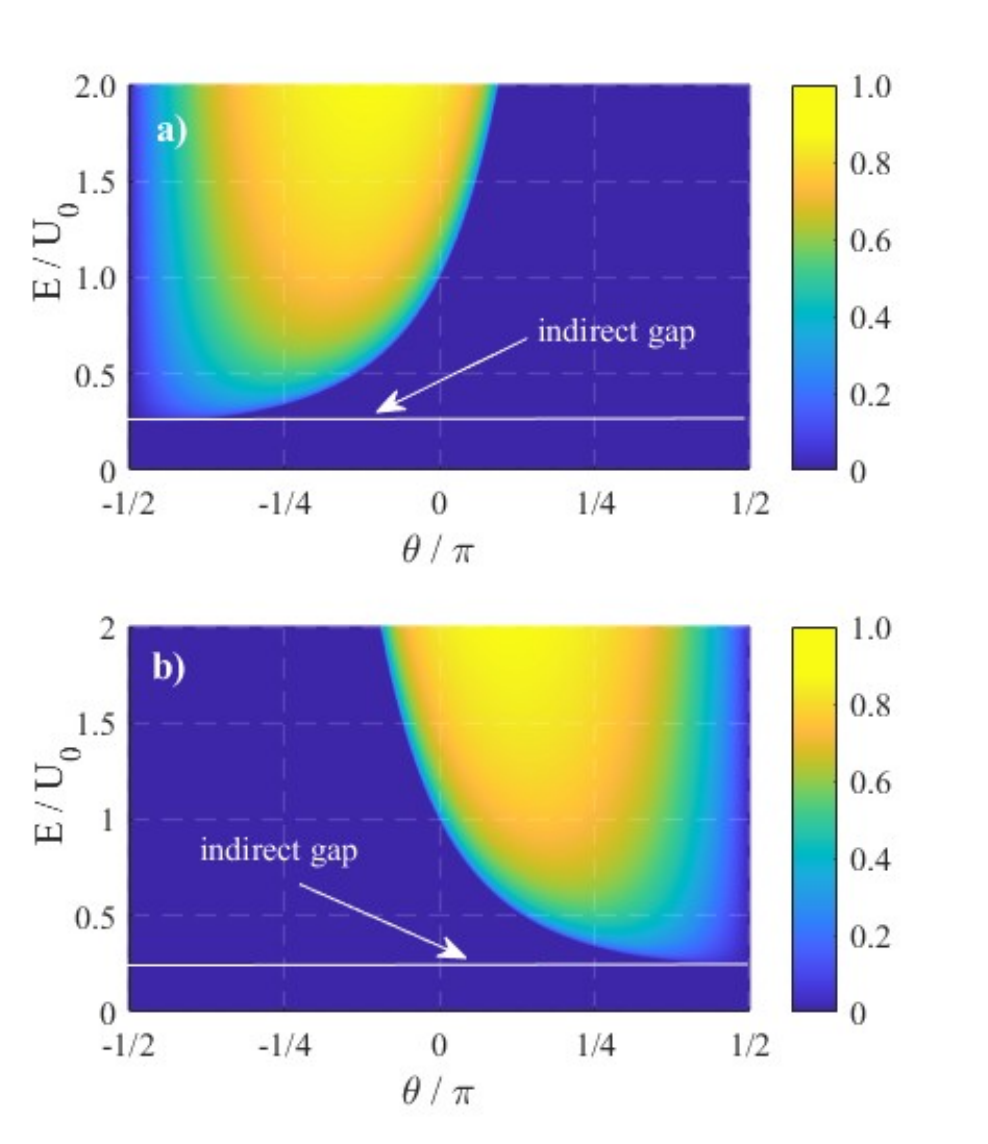}
\caption{a) The anisotropic angle-dependent transmission probability for a spin-$\uparrow$ fermion is given with energy that is normalized by the interface's potential $U_0$. Here, $\theta$ is the propagation angle of the incoming fermion. There is an indirect gap, $\Delta=\alpha_y^2/4$, where all incoming fermions are reflected. In part (b) The transmission probability for spin-$\downarrow$ fermions occurs inversely with respect to perpendicular direction, $\theta=0$. We assume $\alpha_y= U_0$ and $\hbar^2/2m=1$. }
\label{Fig.5}
\end{figure}

\subsection{Anisotropic angle-dependent transmission}
The Eq.(\ref{Eq.T_up}) demonstrates that the interface of the normal/p-wave magnet junction is sensitive to the propagation direction of incoming and outgoing fermions. In a similar way, one can obtain the transmission probability for spin-$\downarrow$ incoming fermion as follows,

\begin{equation}
T_\downarrow=t_\downarrow t^*_\downarrow=\frac{4 cos\theta \cos\theta'_\downarrow}{(\cos\theta+\cos\theta'_\downarrow)^2+U_0^2}
\label{Eq.T_down}
\end{equation}
In the presence of $\alpha_y \neq 0$, the Fermi circles adjust along the $k_y$-direction. Also, the radius of the Fermi circles is tuned by the energy of fermions. To clarify further, we set $\hbar^2/2m=1$. In the same sub-bands, the transmission becomes non-zero when the Fermi circles of different sides of the junction have a non-zero overlap. Thus, an indirect gap appears with the scale of $\Delta=\alpha_y^2/4$ in the transmission plots. Using the condition of Eq.(\ref{Eq.Ballistic_Conditions}), one can calculate the propagation direction of outgoing fermions. 
When $\alpha_x \neq 0$ moves the Fermi circles along the $k_x$-direction and does not change the Fermi circle's overlap. Therefor, it does not affect the transmission processes.
The probabilities of both spin sub-bands are plotted in Fig.(\ref{Fig.5}). Here, we set $\{\alpha_x=0, \alpha_y \neq 0\}$. The energy of incoming fermions normalized by the interface's potential. These plots show the sensitivity of the interface's junction to the propagation direction of incoming fermions. The spin configuration is another important factor. The mirror plane that creates the mirror symmetry, $\mathcal{M}$-symmetry, locates in $\theta=0$, between the probabilities of different sub-bands. The $\mathcal{M}$-symmetry creates the spin current that flows parallel to the interface. Because of $\mathcal{M}$-symmetry the charge current that flows along the $x$-direction is un-polarized. In the high energy limit, $U_0 \ll E$, where the effect of the indirect gap and Fermi circle's displacements are negligible, the transmission probabilities grow, $ T \sim 1$ and the transverse spin current disappears.

\subsection{Transverse spin current}
\label{Sec.TSC}
As shown in part (a) of Fig(\ref{Fig.junction}), a differential electric potential enforces the fermions and creates the current. In the low temperature limit, $\tau\rightarrow 0$, the charge current is equal to the charge conductance that flows along the $x$-direction. Using the probabilities of Eq.(\ref{Eq.T_up}) and Eq.(\ref{Eq.T_down}), one can obtain the charge conductance that flows across the normal/p-wave magnet junctions\cite{Landauer},

\begin{equation}
\frac{G_q(E)}{G^0_q}= \int_{-\frac{\pi}{2}}^{\frac{\pi}{2}}\left(T_\uparrow(E,\theta)\cos\theta_\uparrow+T_\downarrow(E,\theta)\cos\theta_\downarrow\right)d\theta
\label{Eq.conductance}
\end{equation}

\begin{figure}
\includegraphics*[scale=0.45]{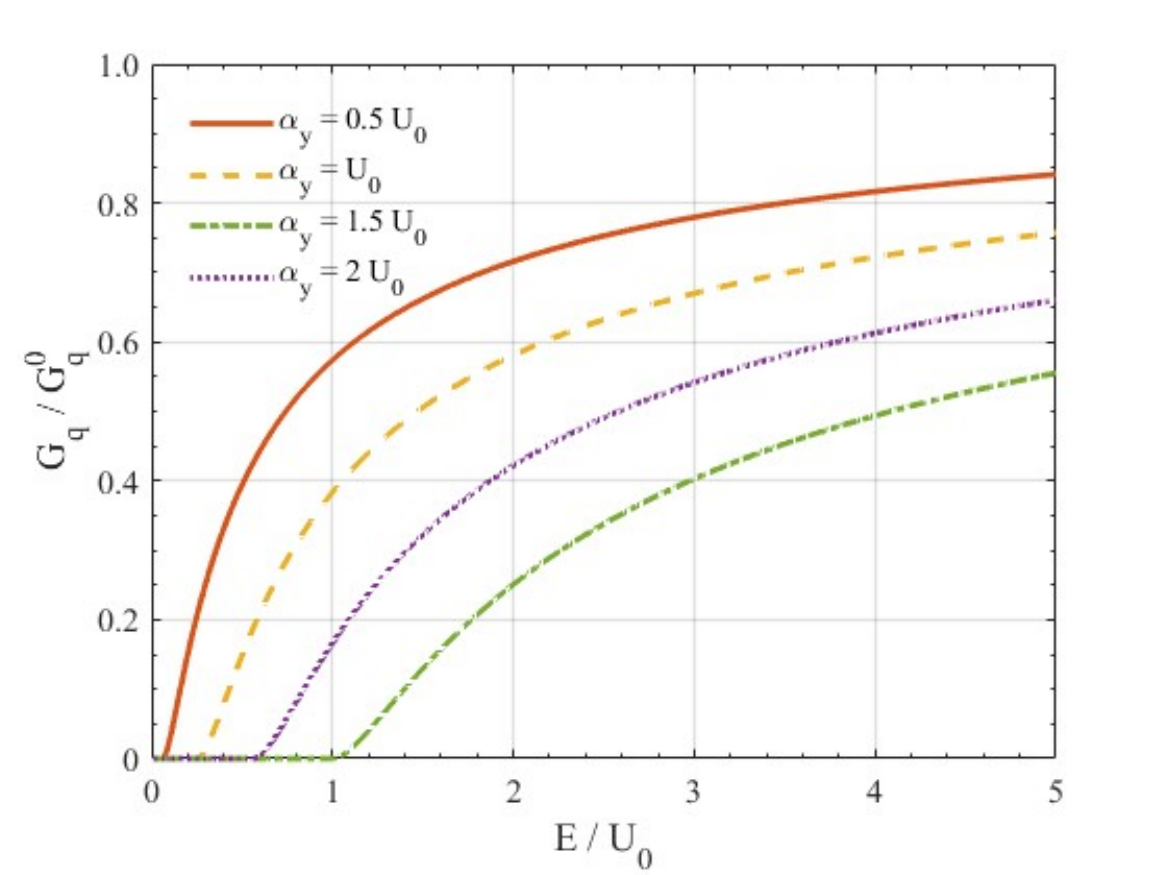}
\caption{ The charge conductance that flows along the $x$-direction versus the energy of excitations. The energy is normalized by the interface's potential, $U_0$. Here, we set $\hbar^2/2m = 1$ and $\mu$=0. The $\boldsymbol{M}$-symmetry creates the un-polarized charge conductance. }
\label{Fig.charge_conductance}
\end{figure}

Here, $G^0_q=e^2/h$ is the quantum of conductance. For different values of $\alpha_y$, the charge conductances are plotted in Fig.(\ref{Fig.charge_conductance}). The conductances start to grow at the edge of the indirect gap monotonically and in high energy values where the Fermi circles displacements and the interface's potential are negligible, tend to perfect value.
Due to the $\mathcal{M}$-symmetry, we have $T_\uparrow(E,\theta)=T_\downarrow(E,-\theta)$ and $\theta_\uparrow=-\theta_\downarrow$. This leads to the zero transverse charge conductance.

On the other hand, the transverse spin conductance is non-zero and can be obtained as follows,
\begin{equation}
\frac{G_s(E)}{G^0_s}=\int_{-\frac{\pi}{2}}^{\frac{\pi}{2}}\left(T_\uparrow(E,\theta)\sin\theta_\uparrow-T_\downarrow(E,\theta)\sin\theta_\downarrow\right)d\theta
\end{equation}

\begin{figure}
\includegraphics*[scale=0.34]{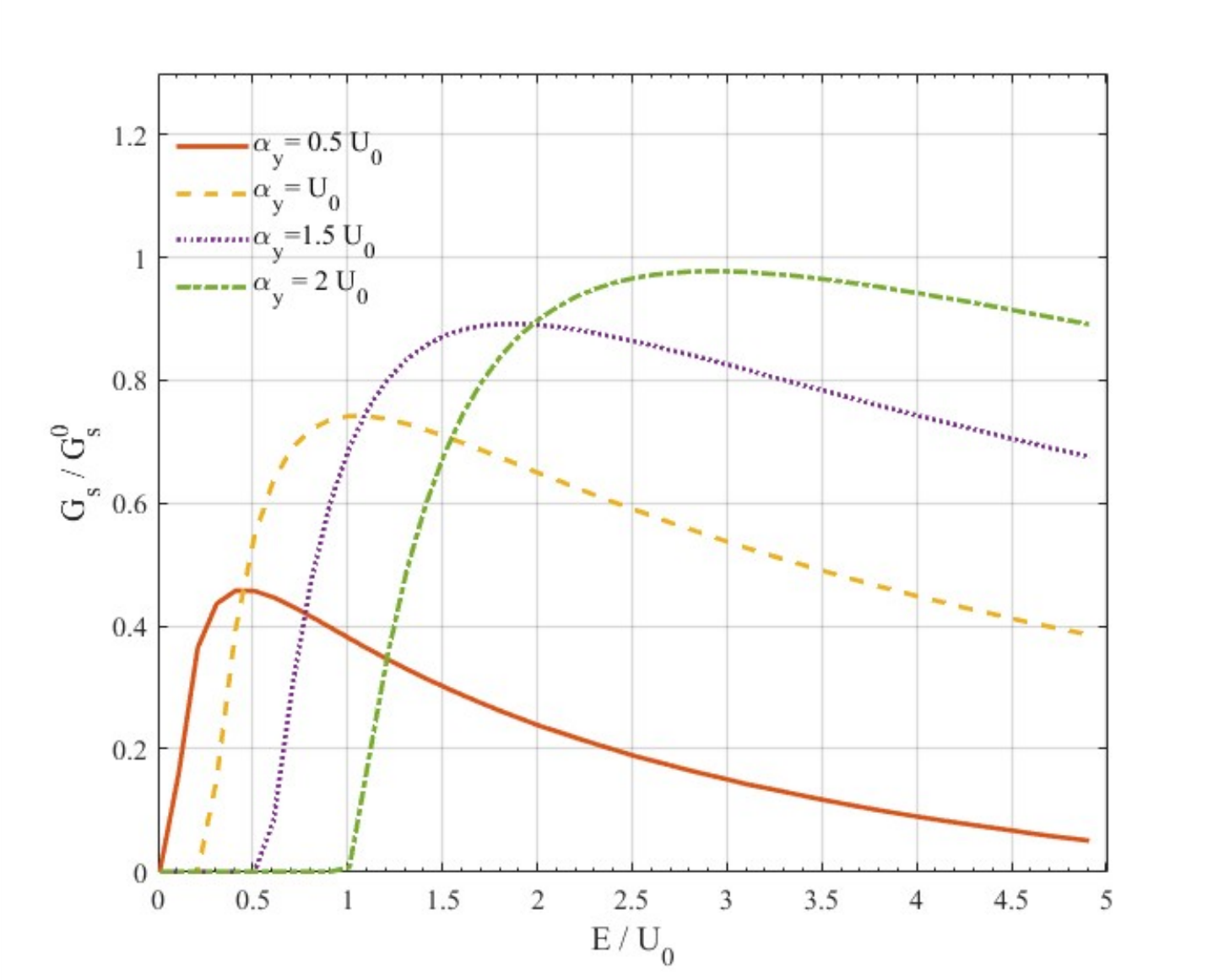}
\caption{ The transverse spin current versus the energy of excitations. The energy is normalized by the interface's potential, $U_0$. Here, we set $\hbar^2/2m = 1$. Due to the existence of the indirect gap, the spin current is zero in the energy range of $0\leq E\leq \Delta$. There is a maximum in each curve around $\sim 2 \Delta$. In the high energy regime the spin current tends to zero. }
\label{Fig.Spin_conductance}
\end{figure}

Here, $G^0_s=e/4\pi$ is the quantum of spin conductance\cite{Kane2005PRL}. For different values of the magnet strength vector, the transverse spin conductances are plotted in Fig(\ref{Fig.Spin_conductance}). We set $\alpha_x=0$, since it does not change the Fermi circle's overlap region. According to  Eq.(\ref{Fig.4}), the radius of the Fermi circles is related with its energy, $\sim \sqrt{E}$. Due to $T_\uparrow = T_\downarrow = 0$, the transverse spin conductance is zero in the $0\leq E \leq \Delta$. As the energy increases, the Fermi circles overlap and the transverse spin conductance appears. All curves have a maximum value between $2\Delta$ and $3\Delta$ corresponding to the strength of the interface's potential. In the absence of the interface's potential, the transverse spin conductance reaches to the perfect values. In the high energy regime, where the displacement of the Fermi circle is negligible, the transverse spin conductance tends to zero. Since the creation of spin current is one of the main issue in spintronics, the p-wave magnets can be good candidates in this regard.

\subsection{ the steady-state approximation}
The role of p-wave magnet strength vector in the creation of transverse spin current can be clarified in the steady state approximation of the Eq.(\ref{Eq.Continuity_Sz}).
In this approximation, $\partial_t \hat{S}_z =0$, we have,

\begin{equation}
\int_{\nu}^{} \nabla_\textbf{r}.\textbf{J}^s(\textbf{r})d\nu=-\frac{\boldsymbol{\alpha}}{m}.\int_{\nu}^{}\nabla_\textbf{r}\rho(\textbf{r})d\nu.
\label{Eq.Steady1}
\end{equation}
Here, $\nu$ stands for the volume integration over the region of interest. By considering a square with the $W$ length on the normal/p-wave magnet junction, we can proceed further. Using the divergence theorem, the volume integration converts to the surface    one as,

\begin{equation}
\alpha_i=\frac{-m}{W \delta N_i}\oint_\mathcal{A} \textbf{J}^s(\textbf{r}).d\boldsymbol{\mathcal{A}}
\label{Eq.Steady2},
\end{equation}
where $\boldsymbol{A}$ is the surface that surrounds the volume of $\nu$ and $i=\{x,y \}$ is the direction's index. The $\delta N_i$ is equal to the number of particles that their projection bend during the scattering processes. The Eq.(\ref{Eq.Steady2}) manifests the role of magnet strength vector in the creation of spin-polarized current over the region of interests. The spin-transfer torque takes this role on the ferromagnetic materials. Also, the Eq.(\ref{Eq.Steady2}) shows that the spin-polarized current flows in the direction of non-zero component of magnet strength vector.

\section{Conclusion}
In this work, we construct a two-dimensional tight-binding square lattice that satisfy the conditions of p-wave magnets. We show in the presence of $\boldsymbol{\alpha}\neq 0$, the spin-polarized bands shift horizontally in the $k$-space and break the $\mathcal{P}$-symmetry. The interface of normal/p-wave magnet becomes sensitive to the spin and propagation angle of incoming fermions. We illustrate that there is a $\mathcal{M}$-symmetry between the transmission probabilities of different spin sub-bands. It leads the un-polarized charge conductance that flows accros the junction whereas creates a transverse spin current. Also, we investigate the role of $\boldsymbol{\alpha}$-vector on the dynamics of spin density wave. We show it acts as source for the continuity equation of spin density wave. Our findings are useful in designing junctions that contains p-wave magnets.

\section{Acknowledgement }
We sincerely thank Razieh Beiranvand for the fruitful discussions that greatly contributed to the progress of this work. We also extend our gratitude to the Educational Affairs Department of the Bu-Ali Sina university for providing the supportive environment and resources necessary for the completion of this research.

\end{document}